\documentclass[cernpreprint,texlive=2016,UKenglish,texmf]{atlasdoc}
 
\usepackage{atlaspackage}
\usepackage{atlasbiblatex}
 
\usepackage{atlasphysics}
\usepackage{siunitx}
 
\graphicspath{{logos/}{figures/}}
 
\usepackage{ANA-STDM-2017-21-PAPER-defs}

\AtlasTitle{Observation of photon-induced $W^+W^-$ production in $pp$ collisions at $\sqrt{s}=13$ TeV using the ATLAS detector}
 
\AtlasAbstract{
This letter reports the observation of photon-induced production of $W$-boson pairs, $\gamma\gamma\rightarrow WW$. The analysis uses 139~fb$^{-1}$ of LHC proton--proton collision data taken at $\sqrt{s}=13$~TeV recorded by the ATLAS experiment during the years 2015--2018. The measurement is performed selecting one electron and one muon, corresponding to the decay of the diboson system as  $WW\rightarrow e^\pm\nu\mu^\mp\nu$ final state. The background-only hypothesis is rejected with a significance of well above 5 
standard deviations consistent with the expectation from Monte Carlo simulation. A cross section for the $\gamma\gamma\rightarrow WW$ process of \ensuremath{3.13\pm0.31\textrm{(stat.)}\pm0.28\textrm{(syst.)}}~fb is measured in a fiducial volume close to the acceptance of the detector, by requiring an electron and a muon of opposite signs with large dilepton transverse momentum and exactly zero additional charged particles. This is found to be in agreement with the Standard Model prediction.
}
 
\author{The ATLAS Collaboration}
 
\AtlasRefCode{STDM-2017-21}
\PreprintIdNumber{CERN-EP-2020-165}
 
\AtlasJournal{Phys.\ Lett.\ B.}
\AtlasJournalRef{Phys.\ Lett.\ B. 816 (2021) 136190}
\AtlasDOI{10.1016/j.physletb.2021.136190}
\hypersetup{pdftitle={ATLAS document},pdfauthor={The ATLAS Collaboration}}
 
\begin{document}
 
\maketitle
 
\tableofcontents

\section{Introduction}
\label{sec:intro}
\begin{figure}[h!]
\centering
\includegraphics[width=0.3\textwidth]{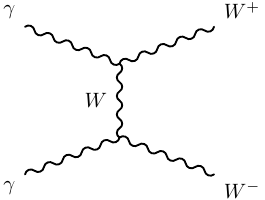}\,\;\;\;\;\hspace{2cm}
\includegraphics[width=0.3\textwidth]{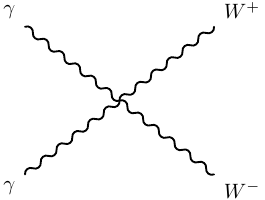}
\caption{
\label{fig:yyWW-feynman-diagrams} The leading-order Feynman diagrams contributing to the \yyWW\ process are the t-channel diagram (left) proceeding via the exchange of a $W$ boson between two $\gamma WW$ vertices and a diagram with a quartic $\gamma\gamma WW$ coupling (right). In addition, a u-channel diagram exists (not shown), which also proceeds via two $\gamma WW$ vertices.}
\end{figure}

The study of $W$-boson pair production from the interaction of incoming photons ($\gamma\gamma \rightarrow WW$) in proton--proton ($pp$) collisions offers a unique window to a wide range of physical phenomena. In the Standard Model (SM), the $\gamma\gamma \rightarrow WW$ process proceeds through trilinear and quartic gauge-boson interactions. This process is unique in that, at leading order, it only involves diagrams with self-couplings of the electroweak gauge bosons, as shown in Figure~\ref{fig:yyWW-feynman-diagrams}. Hence, a cross-section measurement directly tests the SU(2)$\times$U(1) gauge structure of the SM. At the same time, as a process driven only by electroweak boson self-interactions, it is sensitive to anomalous gauge-boson interactions~\cite{Chapon:2009hh} as parameterised in effective field theory (EFT) with additional dimension-6 and dimension-8 operators~\cite{Grzadkowski:2010es,Eboli:2000ad}. Thus, cross-section measurements of \yyWW\ can in future provide valuable input for the global EFT fits.
 
This letter presents a measurement in the $W^+W^-\rightarrow e^\pm\nu\mu^\mp\nu$ channel that results in the observation of photon-induced $WW$ production. Previously, the ATLAS and CMS Collaborations found only evidence for \yyWW\ production with the Run-1 data, ATLAS by using 8~\TeV\ $pp$ collisions~\cite{STDM-2015-10} and CMS by combining their 7~\TeV\ and 8~\TeV\ $pp$ collision data~\cite{CMS-FSQ-12-010,CMS-FSQ-13-008}.
 
The signal process proceeds through the $pp(\gamma\gamma)\rightarrow p^{(*)}W^+W^-p^{(*)}$ reaction, where $p^{(*)}$ indicates that the final-state proton either stays intact or fragments after emitting a photon. Whilst the former occurs through a coherent photon radiation off the whole proton without disintegration, for the latter at least one of the photons can be considered as being radiated off a parton in the proton. These contributions are classified as elastic, single-dissociative, and double-dissociative $WW$ production.
Elastic \yyWW\ production with leptonic decays of the $W$ bosons results in a final state containing two charged leptons and no additional charged-particle activity.
Even in the case of dissociative photon-induced production, the charged particles from the proton remnants often fall outside the acceptance of the tracking detector.
 
The suppressed activity in the central region of the detector in the
\yyWW\ signal gives the means to control and significantly reduce background from quark- and gluon-induced $WW$ production or top-quark production where the leptonic final state is typically produced in association with a substantial amount of hadronic activity. The analysis therefore selects events that have no additional charged-particle tracks reconstructed in the vicinity of the selected interaction vertex.
The modelling of the hadronic activity in quark- and gluon-induced processes, as well as uncorrelated activity from additional $pp$ interactions, is constrained using same-flavour $ee$ and $\mu\mu$ Drell--Yan, DY($\to ee/\mu\mu$), events in data, reducing the associated uncertainties by a significant amount. Background from other photon-induced processes, mainly dilepton production \yyll, is reduced by selecting only different-flavour lepton pairs, $e\mu$, leaving a smaller contribution from \yytt\ production with leptonic $\tau$ decays. Since the contribution from the \yytt\ process falls off rapidly with increasing transverse momentum of the dilepton system, \ptemu, it can be further suppressed by placing requirements on \ptemu.
A fiducial cross section for the $pp(\gamma\gamma)\rightarrow p^{(*)}W^+W^-p^{(*)}$ process through the decay channel $W^+W^-\rightarrow e^\pm\nu\mu^\mp\nu$ is measured in a fit to the number of events in several kinematic regions with different signal and background contributions.
 
\section{ATLAS detector}
\label{sec:detector}
The ATLAS detector~\cite{PERF-2007-01} at the Large Hadron Collider (LHC) is a multipurpose detector with a forward--backward symmetric cylindrical geometry and nearly 4$\pi$ coverage in solid angle.\footnote{ATLAS uses a right-handed coordinate system with its origin at the nominal interaction point in the centre of the detector and the $z$-axis coinciding with the axis of the beam pipe. The $x$-axis points from the interaction point to the centre of the LHC ring, and the $y$-axis points upward. The pseudorapidity is defined in terms of the polar angle $\theta$ as $\eta=-\ln \tan(\theta/2)$, and $\phi$ is the azimuthal angle around the beam pipe relative to the $x$-axis.
The angular distance is defined as $\Delta R=\sqrt{(\Delta \eta)^2+(\Delta \phi)^2}$.
} It consists of an inner tracking detector surrounded by a thin superconducting solenoid providing a \SI{2}{\tesla} axial magnetic field, electromagnetic and hadron calorimeters, and a muon spectrometer.
 
The inner tracking detector (ID) covers the pseudorapidity range \(|\eta| < 2.5\) and is composed of three subdetectors. The high-granularity silicon pixel detector covers the vertex region and typically provides four measurements per track, the first hit normally being in the insertable B-layer~\cite{ATLAS-TDR-2010-19,PIX-2018-001}. It is followed by the silicon microstrip tracker (SCT), which usually provides eight measurements per charged-particle track. These silicon detectors are complemented by the transition radiation tracker, which enables radially extended track reconstruction up to \(|\eta| = 2.0\) and provides electron identification information. The resolution of the $z$-coordinate of tracks at the point of
closest approach to the beam line is about $0.170\,$mm for tracks with $\pt=500$ MeV and improves with higher track momentum
~\cite{ATL-PHYS-PUB-2015-018}. For tracks with $\pt<1$~GeV, the dominant contribution to the $z$-resolution is due to multiple scattering.
 
Lead/liquid-argon (LAr) sampling calorimeters provide electromagnetic (EM) energy measurements with high granularity. A steel/scintillator-tile hadron calorimeter covers the central pseudorapidity range (\(|\eta| < 1.7\)). The endcap and forward regions are instrumented with LAr calorimeters for EM and hadronic energy measurements up to \(|\eta| = 4.9\).
The muon spectrometer (MS) surrounds the calorimeters and is based on
three large air-core toroidal superconducting magnets with eight coils each.
The muon spectrometer includes a system of precision tracking chambers ($|\eta|<2.7$) and fast detectors for triggering ($|\eta|<2.4$).
A two-level trigger system~\cite{TRIG-2016-01} selects the events used in the analysis.
 
\section{Data and simulated event samples}
\label{sec:theory}
The analysis uses proton--proton collision data recorded with the ATLAS detector during the Run-2 data-taking period (2015--2018) at $\sqrt{s}$ = 13~\TeV\ with the number of interactions, $\mu_\textrm{int}$, per bunch crossing (also referred to as pile-up) ranging from about 10 to 60 with an average of 33.7~\cite{ATLAS-CONF-2019-021}.
 
The size of the region where the collisions occur, the so-called beam spot, is a result of the operating parameters of the LHC. Of specific importance for this
analysis is its width along the $z$-direction, which determines the
density of $pp$ interactions. The width is determined by fitting the distribution of the $z$ positions of the reconstructed vertices to Gaussian functions using an unbinned likelihood fit. It varied between 30 and 50 mm during the Run-2 data-taking period~\cite{ATLAS-CONF-2010-027}. The data correspond to an integrated luminosity of $\cal{L} = $139.0~$\pm$~2.4~fb$^{-1}$ after
data quality requirements~\cite{DAPR-2018-01} have been applied.
This value is derived from the calibration of the luminosity scale with the method explained in Ref.~\cite{ATLAS-CONF-2019-021}, using the LUCID-2 detector~\cite{LUCID2} for the primary luminosity measurement.
 
Signal and background processes were modelled using Monte Carlo (MC) event generators to study kinematic distributions, to evaluate background contamination in the signal region and to interpret the results. To simulate the detector response, the generated events were passed through a detailed simulation of the ATLAS detector~\cite{SOFT-2010-01} based
on $\GEANT4$~\cite{Agostinelli:2002hh} or on a combination of $\GEANT4$ and a parameterised calorimeter simulation~\cite{ATL-PHYS-PUB-2010-013}.  The present measurement relies only on tracking information from charged hadrons, muons and electrons, which is simulated by $\GEANT4$ in either case, as well as the modelling of the calorimetric response of electrons which can be reliably parametrized. Multiple $pp$ interactions occurring in the same or adjacent bunch crossings are included in the simulation by overlaying several inelastic $pp$ collisions matching the average number of interactions per bunch crossing. The inelastic $pp$ collisions were generated with \pythia\ {8.186}~\cite{Sjostrand:2014zea} using a set of tuned parameters called the A3 tune~\cite{ATL-PHYS-PUB-2016-017} and the \nnpdftwo~\cite{Ball:2012cx} set of parton distribution functions~(PDF). All MC samples are corrected to the beam conditions of the data as described in Section~\ref{sec:pumodel}. In all samples using \pythia8\ or \Herwig\ to simulate the parton showering, underlying event and hadronisation, the decays of bottom and charm hadrons were performed with EvtGen 1.2.0~\cite{Lange:2001uf}.
 
The elastic component of the \yyWW\ signal process was modelled at leading order (LO) using \HerwigNoSeven\ {7.1.5}~\cite{Bahr:2008pv,Bellm:2015jjp} interfaced with the BudnevQED photon flux~\cite{Budnev:1974de} through \thepeg\ software~\cite{Lonnblad:2009zz}. This sample
is used to model the photon-induced processes in the fiducial region of the measurement as it uses a photon flux, which is differential in both $x$ and virtuality $Q^2$. It is corrected to match the cross section, including the dissociative as well as non-perturbative components, using a data-driven method described in Section~\ref{sec:sigmodel}. This data-driven approach is validated using elastic and dissociative \yyWW\ samples produced using \MGAMCATNO\ {2.6.7}~\cite{Alwall:2014hca} interfaced to \pythia\ {8.243}. The default photon flux in \MGAMCATNO\ and the CT14QED~\cite{Schmidt:2015zda} PDF were used to model the photon radiation from protons and quarks, respectively. The parametrized detector simulation was used in the generation of the \MGAMCATNO\ samples. They are used whenever regions with reconstructed track multiplicities larger than zero are studied.
 
The production of $\gamma\gamma\rightarrow \ell\ell$, with $\ell=e,\mu,\tau$, was modelled in the same way as for the \yyWW\ signal process. Additional generators were used to validate the modelling of the \yyll\ dissociative events. The single-dissociative processes were modelled using \LPAIR\ 4.0~\cite{Vermaseren:1982cz}.
Alternative \yyll\ double-dissociative samples were produced with \pythia\ {8.240} using the {NNPDF3.1NLOluxQED} PDF set~\cite{Bertone_2018}.
Diffractive QCD-processes and $\gamma\gamma\rightarrow 4\ell$ production were produced using \pythia\ {8.244} and \MGAMCATNO\ {2.6.7} interfaced to \pythia\ {8.243} and studied using particle-level information only. The contribution of these processes was found to be negligible in the signal region of the measurement.
 
The dominant background from quark-induced $WW$ production, also referred to as \ppWW, was modelled at next-to-leading-order (NLO) accuracy using the \powhegbox\ {v2}~\cite{Nason:2004rx, Frixione:2007vw, Alioli:2010xd, Melia:2011tj, Nason:2013ydw} generator interfaced to \pythia8\ and alternatively to \Herwig.
The \powhegbox\ {v2} sample employs the \ctten~\cite{CTEQ10} PDF for the matrix element calculation and is interfaced to \pythia\ {8.212} for parton showering and hadronisation employing the parameter values of the \aznlo\ tune~\cite{STDM-2012-23} and the \ctsix~\cite{Pumplin:2002vw} PDF. Samples using a set of variations in the tune parameters (eigentune variations) sensitive to initial- and final-state radiation, as well as further variations related to multiple parton interactions and colour reconnection, were produced to study the description of the parton showers and hadronisation. \HerwigNoSeven\ {7.1.6} was used as an alternative parton shower, using the H7UE tune~\cite{Bellm:2015jjp} and the MMHT2014LO PDF set~\cite{Harland_Lang_2015} for events generated with the \powhegbox~v2 generator.
An alternative sample for quark-induced $WW$ production was generated using the \sherpa~\cite{Hoeche:2009rj,Bothmann:2019yzt} event generator in order to evaluate modelling uncertainties. The \sherpa\ {2.2.2} sample uses matrix elements at NLO accuracy in QCD for up to one additional parton and at LO accuracy for up to three additional parton emissions.
The matrix element calculations were matched and merged with the \sherpa parton shower based on Catani--Seymour dipole factorisation~\cite{Gleisberg:2008fv,Schumann:2007mg} using the MEPS@NLO
prescription~\cite{Hoeche:2011fd,Hoeche:2012yf,Catani:2001cc,Hoeche:2009rj}.
The virtual QCD corrections were provided by the \openloops library~\cite{Buccioni:2019sur,Cascioli:2011va,Denner:2016kdg}. The sample was generated using the \nnpdfnnlo set~\cite{Ball:2014uwa}, along with the dedicated set of tuned parton-shower parameters developed by the \sherpa authors.
 
DY production, $pp\rightarrow Z/\gamma^*\rightarrow \ell\ell$ with $\ell=e,\mu,\tau$, was modelled using the same settings for \sherpa, \powheg+\pythia8\ and \powheg+\Herwig\ as for the quark-induced $WW$ event generation described above. DY($Z/\gamma^*\to \tau\tau$) was modelled with \powheg\ interfaced to \pythia\ {8.186} using the NNPDF3.0NLO PDF set~\cite{Ball:2014uwa} and the \aznlo~tune together with the \ctsix\ PDF set for parton showering and hadronisation.
 
The $WZ$ and $ZZ$ background processes were modelled at NLO using \sherpa\ as well as \powhegbox\ {v2} interfaced to \pythia\ {8.212} with the same settings as employed for the $WW$ event generation.  $W\gamma$ production, gluon-induced $WW$ production including resonant and non-resonant contributions and $WWjj$ production in vector-boson scattering  were simulated using the \sherpa\ {2.2.2} generator with the NNPDF3.0NNLO PDF set. These samples use matrix elements at NLO QCD accuracy for up to one additional parton and LO accuracy for up to three additional parton emissions for $W\gamma$ and gluon-induced $WW$ production and LO-accurate matrix elements for $WWjj$ production in vector-boson scattering.
 
The \ttbar and $Wt$ processes were simulated with the \powhegbox~\cite{Frixione:2007nw,Re:2010bp,Nason:2004rx,Frixione:2007vw,Alioli:2010xd}~v2 generator at NLO with the NNPDF3.0NLO PDF interfaced to \pythia\ {8.230} using the \afourteen\ tune~\cite{ATL-PHYS-PUB-2014-021} and the \nnpdftwo set of PDFs. For the $Wt$ process, the diagram removal scheme~\cite{Frixione:2008yi} was applied to remove interference and overlap with \ttbar\ production.
 
\section{Event reconstruction and selection}
\label{sec:selection}
Candidate events from \yyWW\ production are identified by the presence of an electron and a muon
with high transverse momentum
and the absence of additional reconstructed charged-particle tracks associated with the interaction vertex.
 
Tracks are reconstructed from position measurements (hits) in the ID caused by the passage of charged particles~\cite{ATL-PHYS-PUB-2015-006,ATL-PHYS-PUB-2015-051}. The track reconstruction consists of an iterative track-finding algorithm seeded by combinations of at least three silicon-detector hits followed by a combinatorial Kalman filter~\cite{Fruhwirth:1987fm} to build track candidates based on hits compatible with the extrapolated trajectory. Ambiguities between the track candidates are then resolved and quality criteria are applied to suppress combinations of hits unlikely to originate from a single charged particle. At least one hit in the two innermost layers is required if the extrapolated track crosses the sensitive region of an active sensor module. The number of silicon hits in the pixel and SCT detectors must be larger than 9 for $|\eta|\leq1.65$ or larger than 11 for $|\eta|>1.65$, with no more than two missing SCT hits on a track if the respective SCT modules are operational. Additionally, a selection is imposed on the transverse impact parameter, $|d_0|<1$\,mm, to reject tracks from secondary interactions. Tracks are required to have $\pt>500$~\MeV\ and be within $|\eta|<2.5$. These selection criteria result in an efficiency of 75--80\% depending on the track \pT. The largest source of inefficiency is hadronic interactions with the detector material.
In simulated events, reconstructed tracks can be classified as originating from the hard scatter or from additional $pp$ collisions by matching the hits that contributed to the track fit to the energy deposited by the charged
particle in the $\GEANT4$ simulation. The respective tracks are counted as \ntrkhs\ and \ntrkpu.
 
Electrons are reconstructed from
energy clusters in the electromagnetic calorimeter that are matched to tracks reconstructed in the ID~\cite{PERF-2017-01,EGAM-2018-01}.
The best-matching track is selected using as criteria track--cluster spatial distance and the number of hits in the silicon detectors~\cite{EGAM-2018-01}. Further tracks may be assigned to the electron candidate if they are likely to originate from interactions with detector material. The pseudorapidity of electrons is required to be within the range of $|\eta| < 2.47$, excluding the transition region between the barrel and endcaps in the LAr calorimeter ($1.37 < |\eta| < 1.52$). Electron candidates are required to have transverse momenta \pt$>20$~\GeV.
 
Muons are built from tracks reconstructed using MS hits matched to ID tracks. A global fit using the
hits from both subdetectors is performed~\cite{PERF-2015-10}. Each muon candidate is matched uniquely to exactly one ID track and is required to satisfy $|\eta| < 2.4$ and \pt$>20$~\GeV.
 
Identification and isolation criteria are applied to electron and muon candidates to suppress non-prompt leptons from hadron decays. Identification criteria are based on shower shapes and track parameters for the electrons, and on track parameters for the muons. The isolation criteria use information about ID tracks and calorimeter energy deposits in a fixed cone of $\Delta R = 0.2$ around each lepton. Electrons must satisfy the `medium' identification criteria as well as the loose isolation criteria described in Ref.~\cite{EGAM-2018-01}, which have a combined efficiency of 75{--}85~\% depending on the electron \pt. Muon candidates are required to satisfy the `medium' identification and loose isolation criteria introduced in Ref.~\cite{PERF-2015-10}, which have an efficiency of about 95\%.
The significance of the transverse impact parameter, defined as the absolute value of $d_0$, divided by its uncertainty, $\sigma_{d_{0}}$, must satisfy $|d_{0}|/\sigma_{d_{0}} < 3$ for muons and $|d_{0}|/\sigma_{d_{0}} < 5$ for electrons.
 
The decision on whether or not to record the event was made by single-electron or single-muon triggers with requirements on lepton identification and isolation similar to those applied offline. The transverse momentum thresholds for these triggers were 24~\GeV\ for electrons~\cite{TRIG-2018-05} and 20~\GeV\ for muons~\cite{TRIG-2018-01} in 2015, whilst during the 2016--2018 data-taking period the thresholds were both raised to 26~\GeV\ and requirements on lepton identification and isolation were tightened. Complementary triggers with higher \pt\ thresholds and no isolation or looser identification criteria were used to increase the
trigger efficiency.
 
Events are required to contain exactly two leptons of opposite electric charge that satisfy the above criteria.
One of the leptons must have transverse momentum exceeding 27~\GeV\ and be matched to an object that provided one of the triggers used for the read-out and storage of the event. The invariant mass of the two selected leptons must exceed $\mll=20$~\GeV. Both same-flavour ($ee/\mu\mu$) and different-flavour ($e\mu$) events are accepted either for auxiliary measurements or for the signal extraction, respectively.
 
The interaction vertex is reconstructed from the two leptons in the event, $\ell_1$ and $\ell_2$, as the weighted average $z$-position of the tracks extrapolated to the beam line:
 
\begin{equation}\nonumber
\zlepvtx = \frac{z_{\ell_1} \sin^2\theta_{\ell_1} + z_{\ell_2}  \sin^2\theta_{\ell_2}}
{\sin^2\theta_{\ell_1} + \sin^2\theta_{\ell_2}},
\end{equation}
 
where $\sin^2\theta_{\ell}$ approximately parameterises the resolution of the $z$-position~\cite{ATL-PHYS-PUB-2015-018}. This definition of the interaction vertex is not biased by the presence of additional tracks from hadronic activity in association with the dilepton pair production or by additional tracks from nearby pile-up interactions. It results in a 30\% higher efficiency than a primary vertex selection based on the sum of
squared track transverse momenta~\cite{ATL-PHYS-PUB-2015-026}. Requirements are placed on each lepton to fulfil $| (z_{\ell} - \zlepvtx) \sin\theta|<0.5$~mm.
 
A window of $\Delta z=\pm 1$~mm around $\zlepvtx$ defines the region in which ID tracks are matched to the interaction vertex. The number of tracks in this window, excluding those used in the reconstruction of leptons, is counted as \ntrk. Signal \yyWW\ event candidates are selected using the exclusivity requirement that $\ntrk=0$. Events with low track multiplicities, $1\leq\ntrk\leq4$, are used to evaluate backgrounds.
The modelling of \ntrk\ is therefore vital to the extraction of the \yyWW\ signal, and this is discussed further in the following section.
 
\section{Modelling of signal and backgrounds}
\label{sec:modelling}
Corrections are applied to the simulated signal and background event samples to adjust the lepton trigger, reconstruction, identification and isolation efficiencies, as well as the energy and momentum resolutions, to those observed in data. The muon momentum scale is corrected in the MC simulation, whilst the electron energy scale is corrected in data~\cite{TRIG-2018-05,TRIG-2018-01,EGAM-2018-01,PERF-2015-10}.
Accurate modelling of the transverse momenta of the bosons is important because of its correlation with the expected charged-particle multiplicity from hadronic activity. The \ptww\ distribution in the MC samples for quark-induced $WW$ production is  reweighted to the theoretical calculation at next-to-next-to-leading-order (NNLO) accuracy in perturbative quantum chromodynamics with resummation of soft gluon emissions up to next-to-next-to-next-to-leading-logarithm (N$^3$LL) accuracy using MATRIX+RadISH~\cite{Kallweit:2020gva,Grazzini:2017mhc,Grazzini:2016ctr,Gehrmann:2014fva,Bizon:2017rah,Monni:2016ktx,Cascioli:2011va,Denner:2016kdg,Gehrmann:2015ora,Catani:2007vq,Catani:2012qa}. A correction for the transverse momentum distribution of dilepton pairs from the DY process is derived from data using $ee$ and $\mu\mu$ final states with an invariant mass within 15~\GeV\ of the nominal $Z$ boson mass corrected for background, and is applied to all DY samples as a function of the generator-level $\ptZ$. Additional data-driven corrections are needed for this analysis to account for (i) mismodelling of the additional $pp$ interactions produced in the same bunch crossing, (ii) mismodelling of the charged-particle multiplicity in the \ppWW\ background process, and (iii) second scatterings and the dissociative contribution to the \yyWW\ signal process.
 
\subsection{Modelling of additional $pp$ interactions}\label{sec:pumodel}
 
Tracks from nearby additional $pp$ interactions can be matched to the  interaction vertex and, thus, lower the efficiency of the exclusivity requirement.
Their number depends on the density of additional $pp$ interactions and the number of tracks originating from these interactions.
Data-driven techniques are used to derive corrections to the simulated events to further improve their description of the data, targeting the density of $pp$ interactions and the number of tracks per interaction separately.
 
The simulated events are reweighted such that the distribution of the average number of $pp$ interactions per bunch crossing reproduces the one measured in the
data.
The longitudinal width of the beam spot, $\sigma^{\text{BS}}$, determines the average density, along $z$, of additional $pp$ interactions near the interaction vertex.
The average longitudinal width of the beam spot varied throughout the data-collection period due to changes in the LHC beam optics. It was about 44~mm in 2015 and between 34 and 38~mm in 2016--2018 compared to 42~mm in MC simulation. The photon-induced MC samples were produced with both, the nominal conditions in MC simulation and also with a beam spot width of 35~mm to study the impact of these settings. Only the latter samples were used in the final analysis. To account for the different densities of additional $pp$ interactions in data and simulation, the beam spot width is effectively corrected by modifying the matching of tracks to the interaction vertex in simulation: tracks classified as originating from a pile-up interaction are counted in \ntrk\ if they have a longitudinal impact parameter $z$ within $1~\textrm{mm} \times\sigma^{\text{BS}}_{\textrm{MC}}/\sigma^{\text{BS}}_{\textrm{Data}}$ of \zlepvtx. The values for $\sigma^{\text{BS}}_{\textrm{Data}}$ are sampled from the LHC run conditions during Run 2 according to the luminosity taken at a given value of $\sigma^{\text{BS}}_{\textrm{Data}}$.
 
An ancillary data measurement is used to determine the correction for the number of tracks from additional $pp$ interactions randomly matched to the interaction vertex, \ntrkpu. In same-flavour $Z\rightarrow \ell\ell$ events, this correction is obtained by counting the number of tracks
satisfying the nominal selection criteria
relative to a random position in $z$ that is well separated from the interaction vertex, $|\zlepvtx - z| > 10$~mm. Each event is sampled multiple times
using non-overlapping regions in $z$. This procedure optimises the statistical power, but does not consider the actual distribution of \zlepvtx\ along $z$. To correct for the resulting bias, \ntrkpu\ is extracted as a function of the $z$-coordinate and weighted with the normalised beam spot distribution.
 
\begin{figure}[htb!!]
\centering
\includegraphics[width=0.49\textwidth]{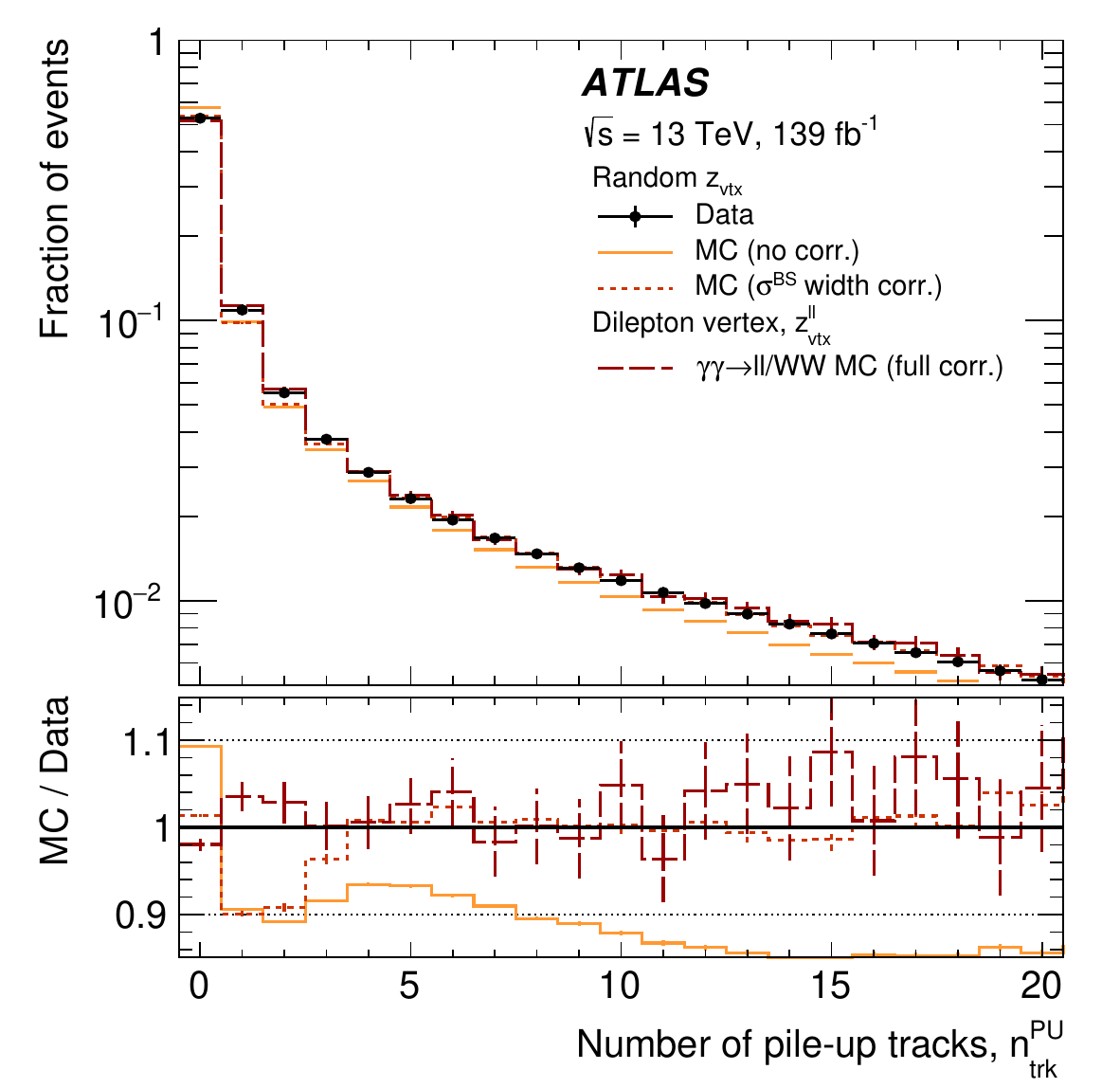}
\caption{The normalised distribution of tracks from additional $pp$ interactions, \ntrkpu, associated with the interaction vertex, in data and signal simulated with a beam spot width of $\sigma^\textrm{BS}_\textrm{MC}=42$~mm. For data, \ntrkpu\ is determined using a random $z$-position along the beam axis away from the interaction vertex. The same quantity is shown for simulated \yyWW\ events before and after correcting the beam spot width to the one observed in data. The inverse ratio of the beam-spot-corrected simulation to data corresponds to the correction applied to \ntrkpu\ in the simulation using the $\GEANT4$-based classification.
To demonstrate the closure of the correction, the number of tracks reconstructed in elastic \yyWW\ signal MC samples is shown after applying the full set of corrections, namely the $\sigma^{\text{BS}}$ correction and the \ntrkpu\ correction. The shown uncertainties are statistical only.
}
\label{fig:PUmodelling}
\end{figure}

This method is tested using simulated events and found to reproduce
the \ntrkpu\ distribution in data within 0.1--3.5\% for low track multiplicities, with larger disagreement for larger \ntrk. Figure~\ref{fig:PUmodelling} shows the probability distribution of \ntrkpu\ associated with \zlepvtx, extracted in data and simulation before and after the corrections for the beam spot width. The bottom panel shows the ratio to data. The inverse ratio of the beam-spot-corrected simulation to data corresponds to the correction applied as a function of \ntrkpu\ in the simulation. The distributions of the number of \ntrkpu\ in \yyll\ and \yyWW\ MC events are shown after the beam spot and the pile-up corrections.
Before any corrections, the disagreement can be up to 15\% depending on the beam spot conditions in the simulation. After the $\sigma^{\text{BS}}$ correction, for low track multiplicities disagreements of about 10\% persists because the $\sigma^{\text{BS}}$ correction only improves the modelling of the density of the pile-up vertices but not of their track multiplicity. This is corrected using the \ntrkpu\ correction. The full set of corrections is applied to all MC samples used in the analysis.
 
The presence of the additional tracks from pile-up will randomly lead to the rejection of signal events and therefore the distribution of $\ntrkpu$ can be used to extract the signal efficiency of the exclusivity requirement ($\ntrk=0$). This exclusive efficiency depends strongly on the number of interactions per bunch crossing and the general beam conditions. The average efficiency for the 2015--2018 dataset with an average $\mu_{\textrm{int}}$ of 33.7 is 52.6\%. It drops from 60\% at $\mu_{\textrm{int}}=20$ to about 30\% at $\mu_{\textrm{int}}=60$. When comparing the data-driven efficiency with that obtained directly from signal MC samples, the results agree to better than 0.2\%.
 
The full effect of the data-driven correction for tracks from additional $pp$ interactions is assigned as a systematic uncertainty, resulting in 1\% and 3\%
uncertainty in the efficiency to
select events without any additional associated tracks ($\ntrk=0$) for signal and background, respectively. The uncertainty of having a low number of tracks associated with the vertex ($1\leq\ntrk\leq4$) is 2\% for photon-induced processes and 10\% for quark- and gluon-induced processes.
 
\subsection{Modelling of the underlying event}\label{sec:uemodel}
 
For quark-induced diboson production, additional charged particles can be produced from initial-state radiation or secondary partonic scatters in the same $pp$ collision, also called the underlying event. However, for low values of the number of charged particles, the \nch\ distribution was found to be not well modelled by many of the phenomenological models implemented in the  generators~\cite{STDM-2017-28,STDM-2016-07,STDM-2014-07,STDM-2011-42}. The underlying event can be assumed to be similar for quark-induced production of different colourless final states if the transverse momenta of these final states are comparable~\cite{STDM-2011-42}. Therefore, the charged-particle multiplicity in \ppWW\ events can be constrained using data measurements of DY production of $\ell\ell$ pairs in $pp$ collisions. Specifically, the charged-particle multiplicity is measured for $Z\rightarrow \ell\ell$ produced in slices of \ptll. This two-dimensional measurement is then used to correct the DY and diboson simulation. The
general validity of this approach has been tested using DY and diboson samples generated with \powheg+\Pythia8, \sherpa\ and \powheg+\Herwig. The multiplicity spectra of charged particles are found to be very different in the different MC samples, yet relatively similar between the respective DY and diboson processes at a constant value of the boson or diboson \pt\ with the agreement being of the order to 10-20\%.
 
The $Z\rightarrow \ell\ell$ events are selected using the criteria described in Section~\ref{sec:selection} with an additional requirement on the dilepton mass (70~\GeV\ $<\mll<$ 105~\GeV) to suppress contributions from background processes. The contribution of pile-up tracks is estimated from data by
sampling random $z$-positions well separated from the dilepton vertex as discussed in Section~\ref{sec:pumodel}. The background at low track multiplicities is dominated by \yyll\ events, which have a different \ptll\ dependence than DY events and amount to about 5\% of the total events selected with 70~\GeV\ $<\mll<$ 105~\GeV\ and $\ntrk=0$ while their contribution is 0.5\% or smaller for higher track multiplicities. The relative normalisations for the elastic, single-dissociative and double-dissociative \yyll\ as well as the DY process are determined in a fit to the measured \ptll\ distribution in a $\mll>105$~\GeV\ sideband, requiring $\ntrk=0$ and using the shapes from MC simulation. In this sideband, the \yyll\ process contributes about 60\% to the total event sample. The contribution from the \yyWW\ process with a same-flavour final state amounts to less than 1\% of the \yyll\ processes in this kinematic region and is neglected.
The overall normalisations of the different \yyll\ contributions relative to the prediction are compatible within the statistical uncertainty with those from earlier ATLAS studies~\cite{STDM-2016-13}.
 
\begin{figure}[htb!!]
\centering
\includegraphics[width=0.49\textwidth]{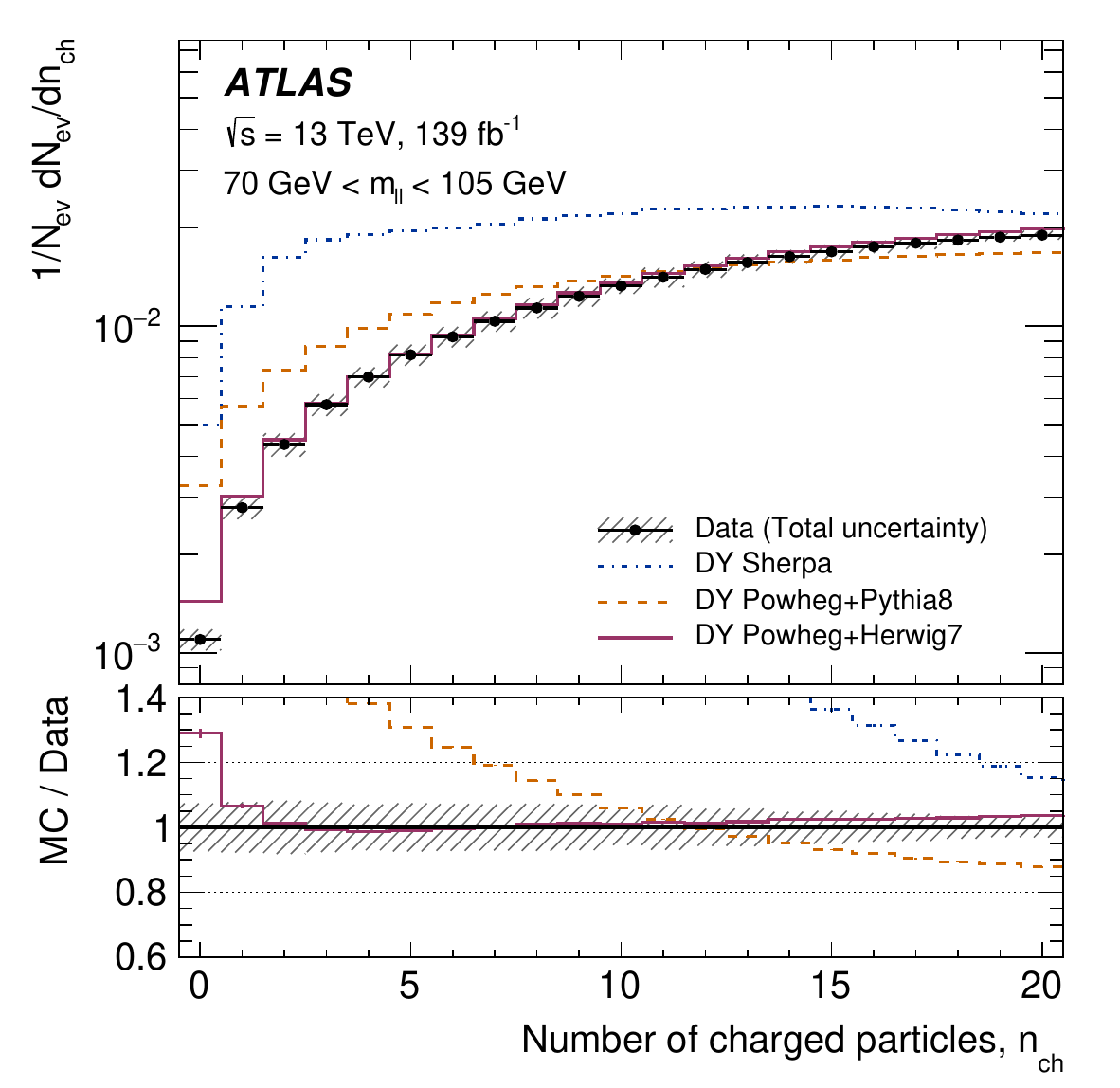}
\includegraphics[width=0.49\textwidth]{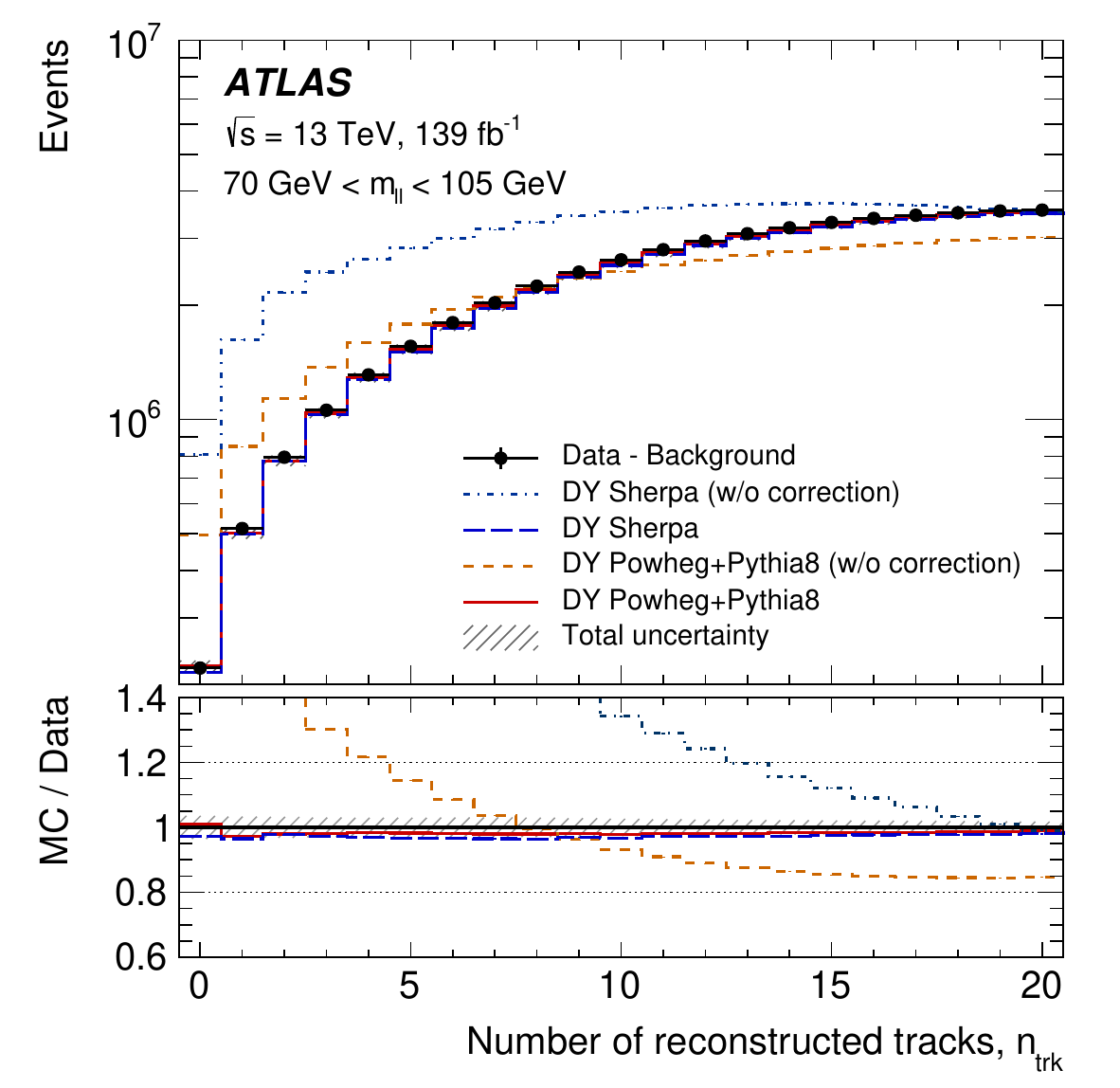}
\caption{On the left, the normalised number of events
with a given number of charged particles, 1/N$_\textrm{ev}$dN$_\textrm{ev}$/dn$_\textrm{ch}$, predicted by \sherpa\, \powheg+\pythia8, and \powheg+\Herwig\ is compared with the unfolded data. The ratio on the bottom is the inverse of the weights that are applied at particle level as a function of the number of charged particles.
The effect of the correction for the underlying event is illustrated for the number of reconstructed tracks on the right. \sherpa and \powheg+\pythia8\ are shown before and after the correction and compared with data.
The total uncertainty of the correction is shown for \powheg+\pythia8\ in the upper panel, and as a band around unity for the lower panel. The total uncertainties for \sherpa\  and \powheg+\pythia8\ are very similar.
}
\label{fig:Nchmodelling}
\end{figure}

After the \yyll\ and pile-up contributions are subtracted as backgrounds, D'Agostini unfolding~\cite{DAgostini:1995,BayesUnfold2} is used to unfold the distribution of the reconstructed track multiplicity, \ntrk, to that of the number of charged particles, \nch, using four iterations.\footnote{Similarly to Ref.~\cite{STDM-2015-02}, charged particles are defined to be stable if they have a mean lifetime $\tau > 30$~ps and satisfy $\pt>500$~\MeV\ and $|\eta|<2.5$.}
The charged-particle multiplicity is extracted as a function of the \pt\ of the dilepton system, which corresponds to the transverse momentum of the recoil, using 5-\GeV-wide intervals of \pt.
The largest sources of uncertainty are the contributions from pile-up tracks and uncertainties in the distribution used as the prior, assessed by comparing \powheg+\pythia8\ and \sherpa.
Other uncertainties originate from the event selection and the \yyll\ background subtraction, assessed by varying the kinematic selection and the normalisation of the photon-induced background within the uncertainties of the fit in the \mll\ sideband. Figure~\ref{fig:Nchmodelling} (left) compares the unfolded charged-particle multiplicity distribution for different MC models and data. For low values of $\nch$, the charged-particle multiplicity distribution is mismodelled by a factor of 2.5 in \powheg+\pythia8\ and by a factor of 4 in \sherpa, whilst good agreement with the \powheg+\Herwig\ model is found except at $\nch=0$ where the \powheg+\Herwig\ prediction exceeds the data yield by about 30\%.
 
The charged-particle multiplicity in simulated DY events is corrected using per-event weights determined as the ratio of the unfolded data to the unfolded MC simulation as a function of the charged-particle multiplicity, and of the particle-level
\pT\ of the decay products of the $Z$ boson.
The impact of the charged-particle multiplicity correction is shown in Figure~\ref{fig:Nchmodelling} (right) for DY events. The simulation is shown both before and after the correction for pile-up modelling and underlying-event modelling in $Z\rightarrow \ell\ell$ events satisfying 70~\GeV\ $<\mll<$ 105~\GeV. The corrections bring the MC simulation into agreement with data within the systematic uncertainty of the charged-particle measurement.
The correction for the underlying-event modelling is applied to $WW$, $WZ$ and $ZZ$ processes as a function of the charged-particle multiplicity, and of the particle-level \pT\ of the decay products of the diboson system.

\subsection{Signal modelling}\label{sec:sigmodel}
 
After the initial \yyWW\ process, the protons can undergo a second inelastic interaction. These additional rescatterings do not change the kinematics of the \yyWW\ process, but lead to the production of particles such that the cross section of \yyWW\ production without associated tracks is reduced. This effect is not included in the modelling of the signal. The probability that no such additional particles are produced is commonly referred to as the survival factor. In addition, the \yyWW\ signal when applying the exclusivity requirement is modelled by \Herwig, which includes only the elastic component. To obtain a better estimate of the expected signal yield including the dissociative components and to correct for effects from the rescattering of protons, a correction factor is obtained from a \yyll\ control sample in data, following a procedure similar to that applied in Refs.~\cite{STDM-2015-10,CMS-FSQ-13-008} using same-flavour lepton final states. To enhance the purity in \yyll\ production and to mimic the kinematic threshold of \yyWW\ production, the dilepton mass is required to be larger than 160~\GeV. The exclusivity requirement of $\ntrk=0$ is applied. In the region, where the correction factor is extracted, the predicted event yield from the \yyWW\ process with same-flavour final states is approximately 1.5\% of the \yyll\ yield so that the derived correction factor is essentially independent of the \yyWW\ signal process.
 
The background, dominated by DY production, is estimated using a data-driven technique. The shape of the \mll\ distribution for background events is estimated using events with $\ntrk=5$, which is a compromise between small signal contamination and closeness to the signal region. This template is normalised to the $\ntrk=0$ selection using a narrow window around the nominal $Z$ boson mass ($83.5~\GeV <\mll< 98.5~\GeV$) where the contribution from photon-induced processes is small. The \mll\ lineshape in simulated DY events is found to be independent of \ntrk for low multiplicities.
 
When the exclusivity requirement of $\ntrk=0$ is applied, the ratio of the yield from photon-induced processes in data to the MC prediction for the elastic processes is found to be $\SFexcl\pm0.15\,\textrm{(tot.)}$. This agrees with the expectation of 3.55 obtained using the MC prediction. It has been verified that the signal modelling correction does not vary as a function of \ptemu within the boundaries used to extract the signal.
 
\begin{figure}[tb!!]
\centering
\includegraphics[width=0.49\textwidth]{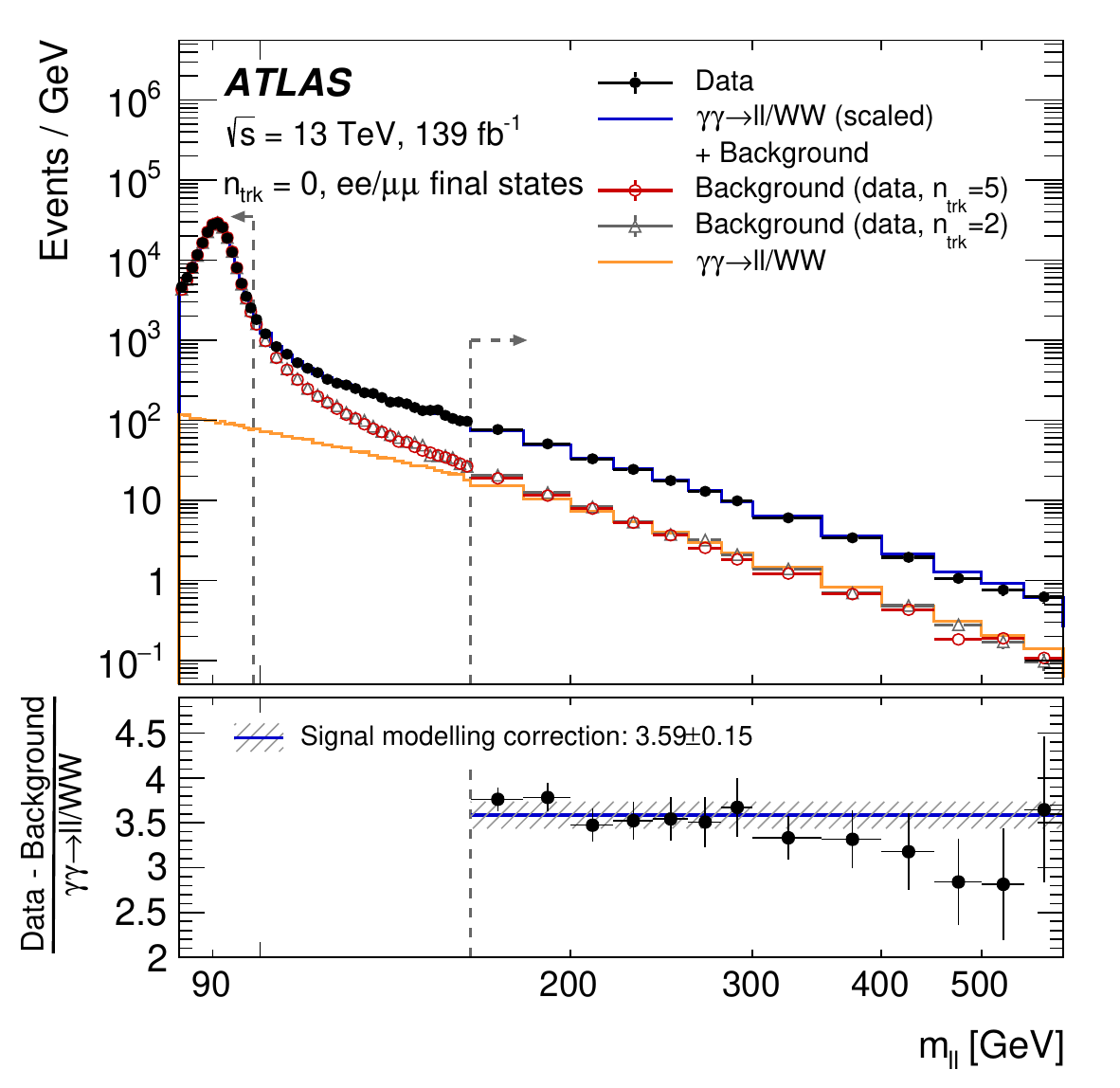}
\caption{The distribution of \mll\ in the region where the signal modelling correction is extracted as the ratio of the yield of \yyll\ and \yyWW\ processes passing the exclusivity requirement of $\ntrk=0$ to the yield of the simulated elastic process only. Shown are the data, where a requirement of $\ntrk=0$ has been applied, and the background templates selected from data using $\ntrk=2$ and $\ntrk=5$. In addition, the \yyll\ and \yyWW\ MC predictions are depicted, as well as the sum of the nominal background template ($\ntrk=5$) and the \yyll\ and \yyWW\ MC predictions scaled by the signal modelling correction.
The normalisation region around the nominal $Z$ boson mass is indicated with a vertical dashed line, as is the region where the signal modelling correction is extracted ($\mll>160$~\GeV). The excess in data relative to the elastic \yyll\ and \yyWW\ prediction is attributed to the dissociative photon-induced processes and used to extract the signal modelling correction that is shown in the lower panel of the plot. The uncertainties shown are statistical only.  }
\label{fig:signalmodel}
\end{figure}

Figure~\ref{fig:signalmodel} illustrates the extraction of the signal modelling correction from data. The signal modelling correction is only applicable to events with $\ntrk=0$. The simulated \Herwig\ events are used in conjunction with the signal modelling correction for predictions of photon-induced processes in events where the $\ntrk=0$ requirement is applied, while the event samples from \MGF\ are used for predictions in regions with larger track multiplicities.
 
Uncertainties are evaluated by increasing the mass window of the DY background normalisation region to $73.5~\GeV <\mll< 108.5~\GeV$ and by changing the number of tracks used in the selection of the template, using $\ntrk=2$ instead of the nominal value. The resulting uncertainty in the signal modelling correction amounts to 4.2\%.
When the signal modelling correction is applied to \yyWW, an additional transfer uncertainty is included to account for potential differences between \yyll\ and \yyWW\ events due to the fact that rescattering effects are mass-dependent. It is calculated as the largest variation that arises from placing different lower bounds on the evaluation region; the lower bound on \mll\ was varied from $\mll=110$~\GeV\ to 400~\GeV\ in intervals of 10~\GeV. The resulting uncertainty amounts to 11\%. This uncertainty affects only the scaling of the \yyWW\ process and thus the measured signal strength and any cross section prediction derived using the signal correction factor, but cancels out in the measurement of the fiducial cross section.

\section{Event categories and background estimation}
\label{sec:bkg}
One signal region and three control regions, enriched in signal and background events respectively, are defined using the dilepton transverse momentum, \ptemu, and the number of additional tracks associated with the interaction vertex, \ntrk.
The signal region is defined by selecting $\ptemu>30$~\GeV\ and $\ntrk=0$. It has an expected purity of 57\% and an expected background contamination from \ppWW\ production of 33\%.
 
Additional kinematic regions with alternative requirements on $\ptemu$ and \ntrk\ are used to control the modelling of background processes. The first control region is defined by $\ptemu<30$~\GeV\ and $1\leq \ntrk \leq 4$ and helps to constrain the DY$(Z/\gamma^*\to \tau\tau)$ normalisation, as this process contributes 75\% of the selected events in this region. It also has non-negligible contributions from \ppWW\ events and non-prompt leptons.
The second control region is defined by $\ptemu>30$~\GeV\ and $1 \leq \ntrk \leq 4$ and is designed to be enriched in \ppWW\ events, with an expected contribution of about 70\% from that process and minor contributions from the DY process and non-prompt lepton events. An additional control region is selected with $\ptemu<30$~\GeV\ and $\ntrk=0$. It brings some additional control for the modelling of backgrounds specific to events with no tracks, however has a signal contamination of the order of 10\%. The boundaries between these regions are chosen such that good signal--background separation is achieved. In addition, the regions used to control the normalisation of the backgrounds are defined to be topologically very similar to the signal region, which helps to minimise uncertainties in extrapolating the normalisation from the control regions to the signal region.
 
Background events from non-prompt leptons contribute about 6\% of the selected signal candidates in the signal region. The primary source of these backgrounds in dilepton events is $W$+jets production where one of the leptons is prompt and the other stems from light-hadron or heavy-flavour decays. Background events from non-prompt leptons are estimated from a control region where exactly one of the leptons must fail to satisfy some of the lepton identification criteria of the nominal event selection.
All other kinematic selection criteria are the same as for the signal selection. The contribution from non-prompt leptons is then estimated by scaling the number of events in the control region by the ratio of the number of non-prompt leptons passing all identification requirements to those failing some of these requirements. This ratio is measured in data selected with one electron and one muon with the same electric charge, and requiring $1\leq\ntrk\leq4$. Contributions from prompt leptons are subtracted using MC simulation. For the extrapolation to the event samples selected with $\ntrk=0$ a dedicated uncertainty is assigned.
\section{Systematic uncertainties}
\label{sec:systematics}
Uncertainties and their correlations are evaluated in each of the signal and control regions.
The uncertainties in the measurement of tracks originate from uncertainties in the inner detector alignment, the reconstruction efficiency, and the probability to incorrectly reconstruct tracks by including hits from noise or from several tracks. The combined uncertainty amounts to 5--7\% of the event yields for DY and \ppWW\ production, whilst for photon-induced processes these uncertainties are $<1$\% in the regions where these processes contribute significantly.
 
Systematic uncertainties in the event yields due to electron and muon reconstruction, including effects from the trigger and reconstruction efficiencies, energy/momentum scale and resolution, and pile-up modelling are 0.5\% and up to 2\% depending on the process, in the signal and control regions, respectively ~\cite{EGAM-2018-01,PERF-2015-10,TRIG-2018-05,TRIG-2018-01}.
 
The uncertainty in the background from non-prompt leptons is dominated by the uncertainty in the measurement of the ratio of non-prompt leptons passing all identification requirements to those failing some, in particular the subtraction of contributions from genuine leptons in the numerator of that ratio. The resulting uncertainty on this background estimation ranges between 50\% and 100\% depending on the region. The statistical uncertainty in the control region for the estimation of background from misidentified leptons is also a significant source of uncertainty.
 
The uncertainties in the correction of pile-up modelling and the underlying event as well as the uncertainty in the signal modelling correction are described in Section~\ref{sec:modelling}.
The correction for the underlying-event modelling in the $WW$, $WZ$ and $ZZ$ processes is derived in bins of \ptll, but applied as a function of diboson \pt, utilising the fact that there are only relatively small differences in charged-particle multiplicity between the DY and diboson processes. Residual differences are evaluated at the particle level and considered as systematic uncertainties. For the largest source of background, the quark-induced $WW$ process, further studies are made. The predicted event yields are compared for \powheg+\pythia8\ and variations of the \pythia8\ parton-shower tunes, and for \powheg+\Herwig\ and \sherpa, with each prediction using its dedicated underlying-event correction. The event yields agree well for $1\leq\ntrk\leq4$, but disagree in the signal region, $\ntrk=0$. The background yield from the quark-induced $WW$ process is estimated as the average of the highest and lowest value of the various predictions, that is the midpoint of the most extreme predictions as no preference for either model can be deduced from the data. The envelope of all predictions is taken as the upper and lower one-standard-deviation boundary, amounting to $\pm$7\% for events selected with $\ntrk=0$, and amounting to less than 1\% for events selected with $1\leq\ntrk\leq4$. The uncertainties in the total quark-induced $WW$ cross section and the shape of the \ptww\ distribution are taken from the MATRIX+RadISH prediction used to reweight the $WW$ samples, amounting to 5--6\%.
 
Because of the specific event selection of the analysis, large uncertainties are applied to minor backgrounds, where the \ntrk\ modelling cannot be easily studied in data: the uncertainty in the $W\gamma$ normalisation is taken to be $\pm$100\%, whereas uncertainties of $\pm$30\% are used for the  normalisation of top-quark production and $WWjj$ production through vector-boson scattering (VBS) as well as gluon-induced resonant and non-resonant $WW$ production. The numbers are informed by the size of the underlying-event correction in DY and $WW$ events and studies on events with forward jets outside the acceptance of the ID. For the smaller background contributions from $WZ$ and $ZZ$  production the uncertainty is assessed by comparing the event yields predicted by \powheg+\pythia8\ with those predicted in \sherpa\ after applying the underlying-event correction described in Section~\ref{sec:uemodel}.
 
The systematic uncertainty in the measured cross section also includes a contribution due to differences in reconstruction efficiency between elastic and dissociative photon-induced processes as well as an uncertainty due to missing spin correlations in \Herwig, which mainly affects the \ptemu\ modelling. These uncertainties are evaluated separately by comparing the reconstruction efficiency of the elastic-only prediction with that including all production mechanisms and by comparing the reconstruction efficiency between \Herwig\ and \MGF. Their combined effect is $\pm$2\%. Uncertainties stemming from the signal modelling correction are applied to the signal prediction and are discussed in detail in Section~\ref{sec:sigmodel}.
\section{Results}
\label{sec:fitting}
The \yyWW\ signal in proton--proton collisions is extracted using a profile likelihood fit of the estimated signal and background event yields to data. The fit uses the integrated event yields in the four kinematic regions introduced in Section~\ref{sec:bkg}, and the $ee+\mu\mu$ events selected as described in Section~\ref{sec:sigmodel}. It maximises the product of Poisson probabilities to produce the observed number of data events, $N_{\text{obs}}$, in each of these regions~\cite{Cranmer:1456844}.
 
The normalisation of the backgrounds from DY and \ppWW\ processes are free parameters in the fit. The expected elastic \yyll\ and \yyWW\ event yields for $\ntrk=0$ are multiplied by the signal modelling correction discussed in Section~\ref{sec:sigmodel}, which is obtained as described within the fit to preserve the experimental correlations correctly. The event yield for the \yyWW\ signal process is also multiplied by a signal strength that is a free parameter in the fit.
Systematic uncertainties are included in the fit as nuisance parameters constrained by Gaussian functions.
The fit can only constrain the sum of the backgrounds, since the background composition is similar in events selected with $\ntrk=0$ and those selected with $1\leq\ntrk\leq4$. Overall, the uncertainty in the sum of their yields is dominated by the systematic uncertainties assigned to events selected with $\ntrk=0$.
In this fit, the background-only hypothesis is expected to be rejected with a significance of \sigmaexp\ standard deviations.
 
Table~\ref{tab:yields} gives an overview of the number of data events compared to background and signal event yields in the different regions after the fit.
The data yield in the signal region is 307, compared with 132 background events predicted by the best-fit result. The normalisations of the $WW$ and the DY background are constrained with the help of the control regions to be \betaWW\ and \betaDY, respectively.
 
By fitting the signal and background event yields in the signal and control regions, the background-only hypothesis is rejected with a significance of \sigmaobs\ standard deviations, assuming that the systematic uncertainties are Gaussian-distributed up to large values. A signal strength of $\muobs$ is measured relative to the yield of elastic \yyWW\ events predicted by \Herwig\ scaled by the signal modelling correction to account for all photon-induced production mechanisms in a phase space with no tracks associated with the interaction vertex. These results constitute the observation of photon-induced $WW$ production in $pp$ collisions, a process for which only evidence with significances of
3.0~$\sigma$~\cite{STDM-2015-10} and 3.6~$\sigma$~\cite{CMS-FSQ-13-008} was previously reported.
 
\begin{table}[tb!]
\begin{center}
\caption{Summary of the data event yields, and the predicted signal and background event yields in the signal region and control regions as obtained after the fit. The uncertainties shown include statistical and systematic components. Because the fit introduces correlations between systematic uncertainties, the uncertainty in the total expected yield is smaller than its components. The leftmost column of values corresponds to the signal region used to measure \yyWW\ in proton--proton collisions.
The numbers for $qq\rightarrow WW$ also contain a small contribution from gluon-induced $WW$ and electroweak $WWjj$ production. The event yields for other backgrounds include contributions from $WZ$ and $ZZ$ diboson production, top-quark production and other gluon-induced processes.
\label{tab:yields} }
\begin{tabular}{l r@{ $\pm$ }r r@{ $\pm$ }r r@{ $\pm$ }r r@{ $\pm$ }r}
\hline
\hline
{}&
\multicolumn{2}{c}{Signal region} &\multicolumn{6}{c}{Control regions}  \\
{$\ntrk$}&
\multicolumn{4}{c}{$\ntrk= 0$}&\multicolumn{4}{c}{$1\leq\ntrk\leq4$}\\
{\ptemu}&
\multicolumn{2}{c}{$>$ 30~\GeV} &\multicolumn{2}{c}{$<$ 30~\GeV} &
\multicolumn{2}{c}{$>$ 30~\GeV} &\multicolumn{2}{c}{$<$ 30~\GeV} \\
\hline
\hline
$\gamma\gamma\rightarrow WW$ & 174\hphantom{.00} & 20\hphantom{.00} & 45\hphantom{.0} & 6\hphantom{.0} & 95\hphantom{.0} & 19\hphantom{.0} & 24 & 5 \\ \hline
$\gamma\gamma\rightarrow \ell\ell$ & 5.5\hphantom{0} & 0.3\hphantom{0} & 39.6 & 1.9 & 5.6 & 1.2 & 32 & 7 \\
Drell--Yan & 4.5\hphantom{0} & 0.9\hphantom{0} & 280\hphantom{.0} & 40\hphantom{.0} & 106\hphantom{.0} & 19\hphantom{.0} & 4700 & 400 \\
$qq\rightarrow WW$ (incl.\ $gg$ and VBS) & 101\hphantom{.00} & 17\hphantom{.00} & 55\hphantom{.0} & 10\hphantom{.0} & 1700\hphantom{.0} & 270\hphantom{.0} & 970 & 150 \\
Non-prompt & 14\hphantom{.00} & 14\hphantom{.00} & 36\hphantom{.0} & 35\hphantom{.0} & 220\hphantom{.0} & 220\hphantom{.0} & 500 & 400 \\
Other backgrounds & 7.1\hphantom{0} & 1.7\hphantom{0} & 1.9 & 0.4 & 311\hphantom{.0} & 76\hphantom{.0} & 81 & 15 \\
\hline
Total & 305\hphantom{.00} & 18\hphantom{.00} & 459\hphantom{.0} & 19\hphantom{.0} & 2460\hphantom{.0} & 60\hphantom{.0} & 6320 & 130 \\
\hline
\hline
Data & \multicolumn{2}{c}{ 307 } & \multicolumn{2}{c}{ 449 } & \multicolumn{2}{c}{ 2458 } & \multicolumn{2}{c}{ 6332 } \\
\hline
\hline
\end{tabular}
\end{center}
\end{table}
 
\begin{figure}[htb!!]
\centering
\includegraphics[width=0.49\textwidth]{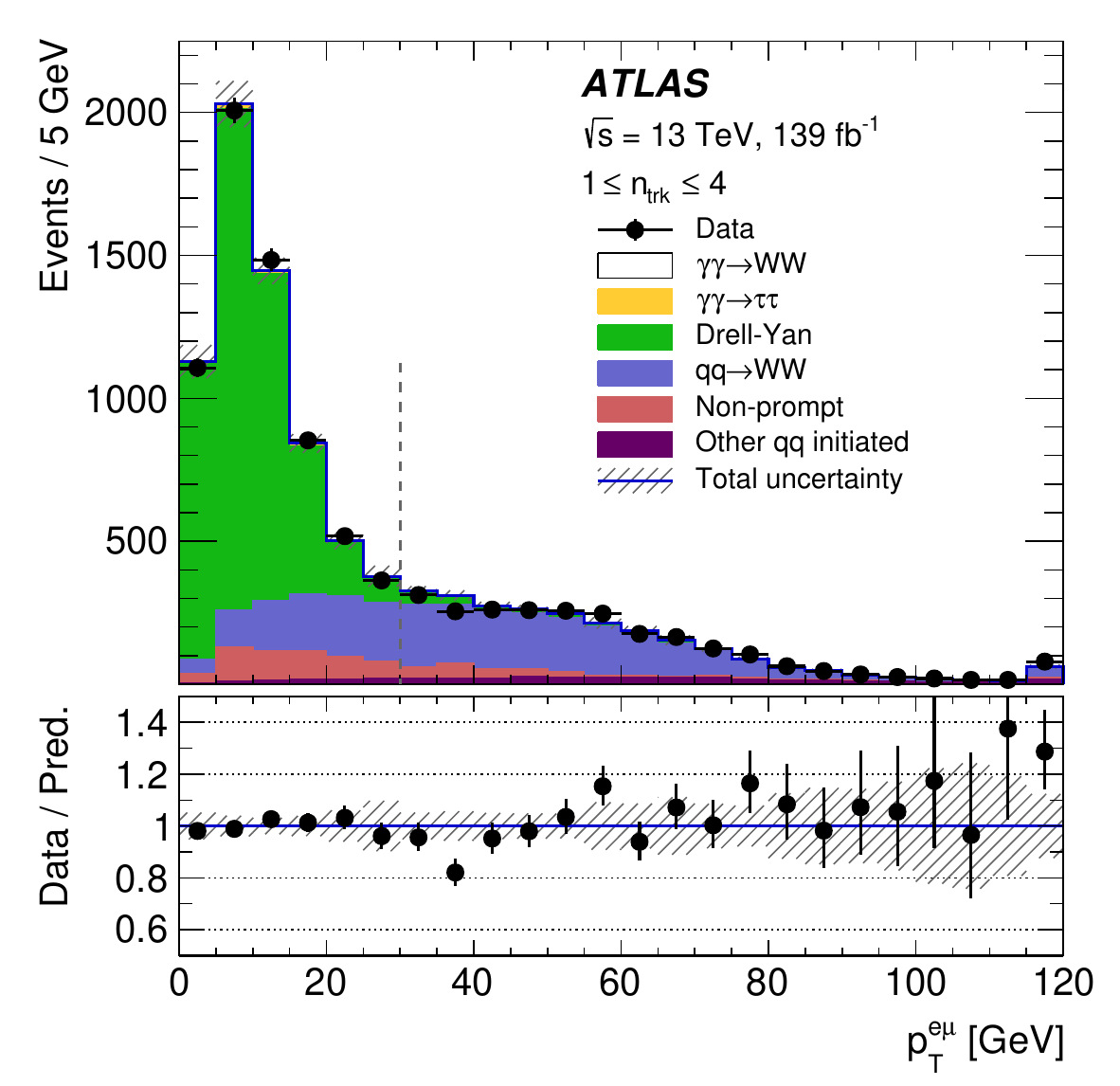}
\includegraphics[width=0.49\textwidth]{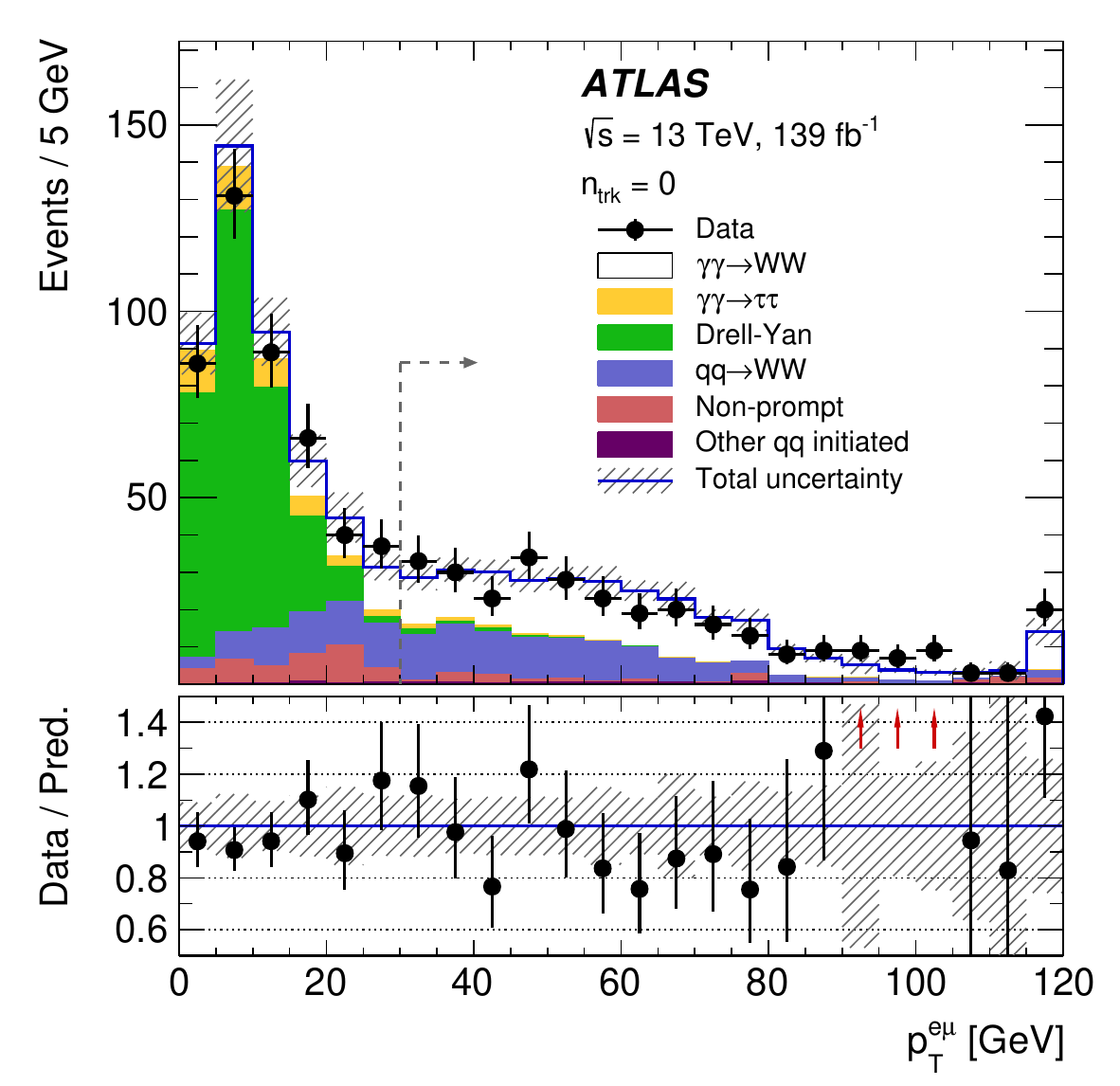}
\caption{The distributions of \ptemu\ for $1\leq\ntrk\leq4$ (left) and $\ntrk=0$ (right) are shown. The fitted normalisation factors and nuisance parameters have been used. The yields for the likelihood fit are given by the integrals of the distributions split at $\ptemu=30$~\GeV, as indicated by the vertical dashed lines. The \yyWW\ signal region requires a selection of $\ptemu>30$~\GeV\ with $\ntrk=0$, as indicated by the arrow.
The $qq\rightarrow WW$ component also contains a small contribution from gluon-induced $WW$ and electroweak $WWjj$ production. Similarly, `other $qq$ initiated' includes contributions not only from $WZ$ and $ZZ$ diboson production but also from top-quark production and other gluon-induced processes.
The total uncertainties are shown as hatched bands.
The lower panels show the ratio of the data to the prediction, with the total uncertainty displayed as a hatched band. An arrow indicates that the ratio is off-scale.
The last bin in both distributions includes the overflow.
}
\label{fig:ptemuplots}
\end{figure}

\begin{figure}[htb!!]
\centering
\includegraphics[width=0.49\textwidth]{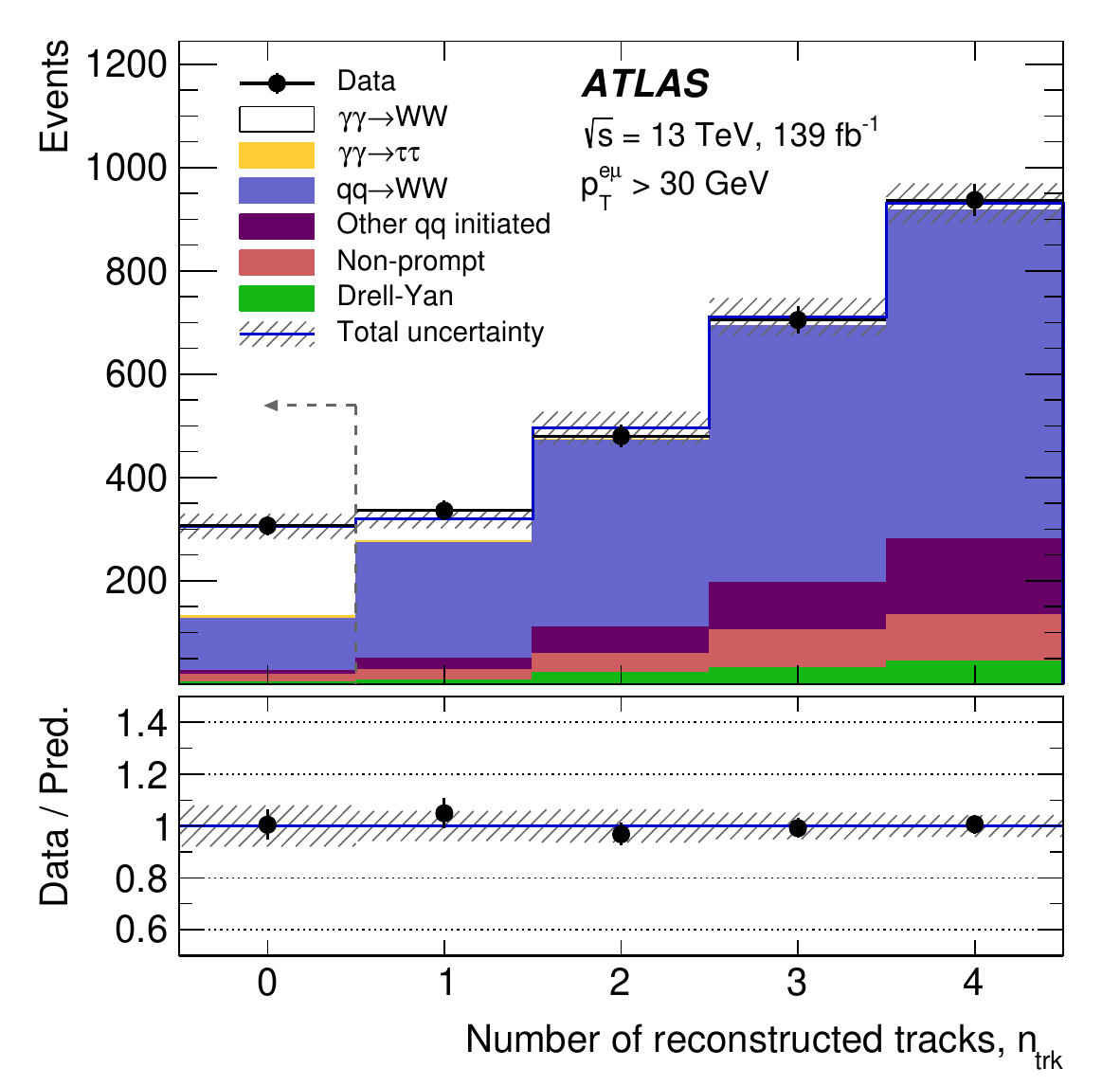}
\caption{The distribution of the number of tracks associated with the interaction vertex is shown. The fitted normalisation factors and nuisance parameters have been used. The \yyWW\ signal region requires a selection of $\ntrk=0$, as indicated by the vertical dashed line.
The $qq\rightarrow WW$ component also contains a small contribution from gluon-induced $WW$ and electroweak $WWjj$ production. Similarly, `other $qq$ initiated' includes contributions not only from $WZ$ and $ZZ$ diboson production but also from top-quark production and other gluon-induced processes.
The total uncertainties are shown as hatched bands.
The lower panel shows the ratio of the data to the prediction, with the total uncertainty displayed as a hatched band.
}
\label{fig:signalmodel_aux}
\end{figure}

Figure~\ref{fig:ptemuplots} shows two \ptemu\ distributions: on the left for events with $1\leq\ntrk\leq4$ associated with the interaction vertex, and, on the right, for events with the exclusivity requirement of no tracks. The boundary between low- and high-\ptemu\ control and signal regions is at 30~\GeV.
The distributions in Figure~\ref{fig:ptemuplots} include the fitted normalisations and nuisance parameters described above; the resulting predictions are in good agreement with the data. Figure~\ref{fig:signalmodel_aux} shows the distribution of the number of reconstructed tracks for $\ptemu>30$~\GeV.
 
The fiducial phase space used for the cross-section measurement is defined to be close to the acceptance of the detector. The leptons must at particle level satisfy the pseudorapidity requirement $|\eta|<2.5$. One of the leptons is required to have a transverse momentum of at least 27~\GeV, whilst the other must have $\pt>20$~\GeV. They are required to be prompt leptons from $W$ decays. Photons in a cone of $\Delta R=0.1$ around a lepton and not originating from the decays of hadrons are added to the four-momentum of the lepton, that is leptons are ``dressed''. Events with exactly two leptons are selected with opposite-sign and different-flavour final states. Decays of either $W$ boson into a $\tau$-lepton and neutrino are excluded. The invariant mass of the dilepton system is required to be $\mll>20$~\GeV\ and its transverse momentum must be \ptemu$> 30$~\GeV. The number of charged particles, \nch, with \pt$>500$~\MeV\ and within $|\eta|<2.5$, excluding the selected leptons, is required to be zero.
 
Without requirements on the number of reconstructed tracks, the selection efficiency after reconstruction is $\cww\%$ for elastic \yyWW\ events in the fiducial region. The full selection efficiency after applying $\ntrk=0$ is $\cwwAll\%$. The predicted number of signal events includes a $\sim$5\% contribution of leptons from $W\rightarrow\tau\nu_\tau$, $\tau\rightarrow\ell\nu_\ell\nu_\tau$, which is estimated using the MC simulation and which is removed from the measured fiducial cross section using this fractional contribution.
 
The observed signal strength translates into a fiducial cross section of \[ \sigma_\textrm{meas}=\xsobs\] for $pp(\gamma\gamma)\rightarrow p^{(*)}W^+W^-p^{(*)}$ production with $W^+W^-\rightarrow e^\pm\nu\mu^\mp\nu$. The uncertainties correspond to the statistical and systematic uncertainties, respectively. Table~\ref{tabaux:systematics} gives an overview of the sources of systematic uncertainties, which are discussed in Section~\ref{sec:systematics} and presents their effect on the measured cross section. To evaluate the impact of one source of systematic uncertainty, the fit is performed with the corresponding nuisance parameter fixed one standard deviation up or down from the value obtained in the nominal fit, then these high and low variations are symmetrised.
 
\begin{table}[htb]
\begin{center}
\caption{The impact of different components of systematic uncertainty on the measured fiducial cross section, without taking into account correlations. The impact of each source of systematic uncertainty is computed by first performing the fit with the corresponding nuisance parameter fixed to one standard deviation up or down from the value obtained in the nominal fit, then these high and low variations are symmetrised. The impacts of several sources of systematic uncertainty are added in quadrature for each component.
\label{tabaux:systematics}}
\begin{tabular}{p{2ex}lc}
\toprule
\multicolumn{2}{l}{Source of uncertainty} & Impact [\% of the fitted cross section] \\
\midrule
\multicolumn{2}{l}{Experimental}           & \\
\midrule 
&Track reconstruction & ${\hphantom{0}1.1}$ \\
&Electron energy scale and resolution, and efficiency  & ${\hphantom{0}0.4}$ \\
&Muon momentum scale and resolution, and efficiency & ${\hphantom{0}0.5}$ \\
&Misidentified leptons, systematic& ${\hphantom{0}1.5}$ \\
&Misidentified leptons, statistical & ${\hphantom{0}5.9}$ \\
&Other background, statistical & ${\hphantom{0}3.2}$ \\
\midrule
\multicolumn{2}{l}{Modelling}           & \\
\midrule
&Pile-up modelling& ${\hphantom{0}1.1}$ \\
&Underlying-event modelling& ${\hphantom{0}1.4}$ \\
&Signal modelling  & ${\hphantom{0}2.1}$ \\
&$WW$ modelling  & ${\hphantom{0}4.0}$ \\
&Other background modelling  & ${\hphantom{0}1.7}$ \\
\midrule
\multicolumn{2}{l}{Luminosity}   & ${\hphantom{0}1.7}$ \\
\midrule
\multicolumn{2}{l}{Total}  & ${\hphantom{0}8.9}$ \\
 
\bottomrule
\end{tabular}
\end{center}
\end{table}
 
The data measurement can be compared with two types of predictions. The first, used in the definition of the signal strength and the calculation of the expected significance, is based on the \Herwig\ prediction for elastic \yyWW\ events scaled by the data-driven signal modelling correction to include the dissociative processes and rescattering effects as described in Section~\ref{sec:sigmodel}. It is found to be \[\sigma_\textrm{theo} \times \left( \SFexcl \pm 0.15~\textrm{(exp.)} \pm 0.39~\textrm{(trans.)} \right) = \xsexp\,,\]
where the uncertainty contains all experimental uncertainties and receives an additional component due to the transfer from the \yyll\ to the \yyWW\ process described above. The uncertainties in the theory prediction are negligible because the scale uncertainty in the calculation of elastic production based on a photon-flux is small and partially cancels with the signal correction that is calculated with respect to the same photon-flux compared to the data. A standalone theory prediction for the fiducial cross section is computed with \MGF\ using the appropriate elastic or inelastic MMHT2015qed PDF sets~\cite{Harland-Lang:2019pla} for each of the contributions by applying the fiducial requirements to all photon-induced contributions, which yields $4.3 \pm 1.0\,\textrm{(scale)} \pm 0.1\,\textrm{(PDF)}$~fb. The scale uncertainty is determined by varying the factorisation scale by factors of 2 and 0.5 and symmetrising the effect. The contributions to this cross-section prediction from elastic and single-dissociative production are 16\% and 81\%, respectively. Double-dissociative production contributes only 3\%. Using CT14qed~\cite{Schmidt:2015zda} as the central PDF set yields a prediction which is 26\% smaller and amounts to $3.2$~fb.
 
The \MGF\ prediction does not include rescattering effects that are expected to
decrease the fiducial cross section.
For elastic \yyWW\ production, a survival factor of 0.65 was estimated in Ref.~\cite{Dyndal_2015}. In Ref.~\cite{Harland-Lang:2015cta} a survival factor of 0.82 was calculated in a two-channel eikonal model also accounting for the helicity structure of the hard scattering process.\footnote{More recent calculations predict slightly lower absorption values for the dissociative processes~\cite{Harland-Lang:2016apc,Harland-Lang:2020veo}.}
Multiplying the \MGF\ prediction by these survival factors results in theoretical predictions of $2.8\pm0.8$~fb and $3.5\pm1.0$~fb, respectively, with the total uncertainties calculated as the quadratic sum of scale and PDF uncertainties. These predictions are in agreement with the measurement.

 
\section{Conclusion}
\label{sec:conclusion}
The photon-induced production process, \yyWW, was studied in proton--proton collisions at $\sqrt{s}=13$~\TeV\ recorded with the ATLAS detector at the LHC
corresponding to an integrated luminosity of \intlumi. Events with leptonic $W$ boson decays into $e^\pm\nu\mu^\mp\nu$ final states
were selected by requiring that no tracks except those of the two charged leptons are associated
with the production vertex. The background-only
hypothesis is rejected with a significance of \sigmaobs\ standard
deviations whereas well above 5 $\sigma$
was expected. This measurement constitutes the observation of photon-induced $WW$ production in $pp$ collisions, a process for which only
evidence was previously reported. The signal strength and the cross section for
the sum of elastic and dissociative production mechanisms are
measured. The cross section for the $pp(\gamma\gamma)\rightarrow p^{(*)}W^+W^-p^{(*)}$ process in the decay channel $W^+W^-\rightarrow e^\pm\nu\mu^\mp\nu$ in a fiducial phase space close to the
experimental acceptance is measured to be \xsobs. This result is in agreement with the theoretical predictions and may serve as input into EFT interpretations.
\FloatBarrier
\section*{Acknowledgements}

We thank CERN for the very successful operation of the LHC, as well as the
support staff from our institutions without whom ATLAS could not be
operated efficiently.
 
We acknowledge the support of ANPCyT, Argentina; YerPhI, Armenia; ARC, Australia; BMWFW and FWF, Austria; ANAS, Azerbaijan; SSTC, Belarus; CNPq and FAPESP, Brazil; NSERC, NRC and CFI, Canada; CERN; ANID, Chile; CAS, MOST and NSFC, China; COLCIENCIAS, Colombia; MSMT CR, MPO CR and VSC CR, Czech Republic; DNRF and DNSRC, Denmark; IN2P3-CNRS and CEA-DRF/IRFU, France; SRNSFG, Georgia; BMBF, HGF and MPG, Germany; GSRT, Greece; RGC and Hong Kong SAR, China; ISF and Benoziyo Center, Israel; INFN, Italy; MEXT and JSPS, Japan; CNRST, Morocco; NWO, Netherlands; RCN, Norway; MNiSW and NCN, Poland; FCT, Portugal; MNE/IFA, Romania; MES of Russia and NRC KI, Russia Federation; JINR; MESTD, Serbia; MSSR, Slovakia; ARRS and MIZ\v{S}, Slovenia; DST/NRF, South Africa; MICINN, Spain; SRC and Wallenberg Foundation, Sweden; SERI, SNSF and Cantons of Bern and Geneva, Switzerland; MOST, Taiwan; TAEK, Turkey; STFC, United Kingdom; DOE and NSF, United States of America. In addition, individual groups and members have received support from BCKDF, CANARIE, Compute Canada and CRC, Canada; ERC, ERDF, Horizon 2020, Marie Sk{\l}odowska-Curie Actions and COST, European Union; Investissements d'Avenir Labex, Investissements d'Avenir Idex and ANR, France; DFG and AvH Foundation, Germany; Herakleitos, Thales and Aristeia programmes co-financed by EU-ESF and the Greek NSRF, Greece; BSF-NSF and GIF, Israel; La Caixa Banking Foundation, CERCA Programme Generalitat de Catalunya and PROMETEO and GenT Programmes Generalitat Valenciana, Spain; G\"{o}ran Gustafssons Stiftelse, Sweden; The Royal Society and Leverhulme Trust, United Kingdom.
 
The crucial computing support from all WLCG partners is acknowledged gratefully, in particular from CERN, the ATLAS Tier-1 facilities at TRIUMF (Canada), NDGF (Denmark, Norway, Sweden), CC-IN2P3 (France), KIT/GridKA (Germany), INFN-CNAF (Italy), NL-T1 (Netherlands), PIC (Spain), ASGC (Taiwan), RAL (UK) and BNL (USA), the Tier-2 facilities worldwide and large non-WLCG resource providers. Major contributors of computing resources are listed in Ref.~\cite{ATL-SOFT-PUB-2020-001}.
 

\clearpage
 
\printbibliography
 
\clearpage
 
\begin{flushleft}
\hypersetup{urlcolor=black}
{\Large The ATLAS Collaboration}

\bigskip

\AtlasOrcid[0000-0002-6665-4934]{G.~Aad}$^\textrm{\scriptsize 102}$,    
\AtlasOrcid[0000-0002-5888-2734]{B.~Abbott}$^\textrm{\scriptsize 128}$,    
\AtlasOrcid[0000-0002-7248-3203]{D.C.~Abbott}$^\textrm{\scriptsize 103}$,    
\AtlasOrcid[0000-0002-2788-3822]{A.~Abed~Abud}$^\textrm{\scriptsize 36}$,    
\AtlasOrcid[0000-0002-1002-1652]{K.~Abeling}$^\textrm{\scriptsize 53}$,    
\AtlasOrcid[0000-0002-2987-4006]{D.K.~Abhayasinghe}$^\textrm{\scriptsize 94}$,    
\AtlasOrcid[0000-0002-8496-9294]{S.H.~Abidi}$^\textrm{\scriptsize 167}$,    
\AtlasOrcid[0000-0002-8279-9324]{O.S.~AbouZeid}$^\textrm{\scriptsize 40}$,    
\AtlasOrcid{N.L.~Abraham}$^\textrm{\scriptsize 156}$,    
\AtlasOrcid[0000-0001-5329-6640]{H.~Abramowicz}$^\textrm{\scriptsize 161}$
% ...
\bigskip
\\

% ...
$^{36}$CERN, Geneva; Switzerland.\\
% ...
$^{40}$Niels Bohr Institute, University of Copenhagen, Copenhagen; Denmark.\\
% ...
$^{53}$II. Physikalisches Institut, Georg-August-Universit\"{a}t G\"ottingen, G\"ottingen; Germany.\\
% ...
$^{94}$Department of Physics, Royal Holloway University of London, Egham; United Kingdom.\\
% ...
$^{102}$CPPM, Aix-Marseille Universit\'e, CNRS/IN2P3, Marseille; France.\\
$^{103}$Department of Physics, University of Massachusetts, Amherst MA; United States of America.\\
% ...
$^{128}$Homer L. Dodge Department of Physics and Astronomy, University of Oklahoma, Norman OK; United States of America.\\
% ...
$^{156}$Department of Physics and Astronomy, University of Sussex, Brighton; United Kingdom.\\
% ...
$^{161}$Raymond and Beverly Sackler School of Physics and Astronomy, Tel Aviv University, Tel Aviv; Israel.\\
% ...
$^{167}$Department of Physics, University of Toronto, Toronto ON; Canada.\\
% ...

\end{flushleft}

 
\end{document}